 \documentclass[11pt]{article}

\usepackage[round]{natbib}   

\usepackage[english]{babel}

\usepackage[utf8]{inputenc} 
\usepackage[T1]{fontenc}
\usepackage{lmodern}

\usepackage{pifont}     

\usepackage{graphicx}
\usepackage{amsmath}

\usepackage{enumitem}

\usepackage{float}
\usepackage{ifthen}
\usepackage{hyperref}

\usepackage{color}

\begin{document}

\begin{center}
{\bf \Large \color{blue} 
The third law of thermodynamics or an absolute definition for
}
\\ 
\vspace*{2mm}
{\bf \Large \color{blue}  
Entropy. Part 1 : the origin and applications in thermodynamics.
}
\\ 
\vspace*{4mm}
{\bf \large \color{black}  
Accepted for the revue: ``\textbf{\emph{La Météorologie}}\,'' / Revision 2 - \today.
}
\\ \vspace*{5mm}
{\Large by Pascal Marquet. M\'et\'eo-France, CNRM/GMAP, Toulouse.}
\end{center}


\begin{center}
{\Large \bf Abstract}

This article describes the third law of thermodynamics. This law is often poorly known and is often decried, or even considered optional and irrelevant to describe weather and climate phenomena. This, however, is inaccurate and contrary to scientific facts. A rather exhaustive historical study is proposed here in order to better understand, in another article to come, why the third principle can be interesting for the atmosphere sciences.

\end{center}

 \section{\underline{\Large Introduction}}
\label{Introduction}
\vspace{-1mm}

Before being able to study in a second part the properties of entropy in the atmosphere, it is necessary to recall in this first part why its calculation poses certain problems in thermodynamics, problems whose solution passes through the invention and the application of the third law of thermodynamics which introduces a kind of absolute in the calculation of entropy.
And the idea that certain absolutes may exist had preceded the establishment of the third law.

 \section{\underline{\Large The notion of absolute temperature}}
\label{section_2}
\vspace{-1mm}

Thermodynamics teaches us that the notion of temperature corresponds to the measurement of the energy of the microscopic agitations of atoms or molecules in the solids, liquids or gases which constitute the environment which surrounds us, and therefore in particular in the atmosphere.

Carnot (1824) was able to establish the existence of universal things, supposing that the perpetual motion of thermal machines was impossible.
He first highlighted the existence of a maximum efficiency that depends only on the temperatures of the bodies between which these machines operate.
He has also shown that the difference of specific heats under constant pressure and volume (the perfect gas constant) is independent of the nature of the bodies studied.
He also gathered for the first time the two laws of Mariotte and Gay-Lussac in a single law which he expressed, in a note on page 67, by the equation:
\vspace{-1mm}
\begin{align}  
 P \: v & \: = \; c  \: \left( t \: + \: 267 \right) \: .
 \label{eq_1}
\end{align}

This equation reflects the fact that the product of the pressure ($P$) by the volume ($v$) is proportional to the temperature, the coefficient of proportionality being a constant denoted ``$c$''.
The temperature is noted here ``($t+267$)'', where ``$t$'' is the temperature in degrees centigrade and where ``$267$'' is an approximation of the inverse of the compressibility coefficient previously measured by Gay-Lussac (1802, page 166).
We therefore recognize in the equation (\ref{eq_1}) the perfect gas law, where nowadays the constant ``$c$'' writes ``$n \: R$'', with ``$n$'' representing the number of moles and ``$R$'' the perfect gas constant $8.314$~J/kg/mol.

But how did we arrive at this definition of Carnot, with this number $267$, to that of a temperature called ``absolute'', where ``$ t + 267 $'' is now replaced by the absolute temperature ``$T = t + 273.15$''?

Carnot's work initially had no impact on the scientific community. 
Things changed with Clapeyron's (1834) writing of a memoir in which he made Carnot's works more understandable, by making more use of mathematical language and using (page 164) the letter ``$R$'' instead of ``$c$'' in (\ref{eq_1}).
It is then the English and German translations of this article by Clapeyron that will allow Carnot's ideas to strongly influence the creation of thermodynamic science in England and Germany in the 19th century.

This was the case in England for the work of William Thomson (the future Lord Kelvin), who first knew the work of Carnot via the translation of Clapeyron's article in 1837, before receiving a copy of the original memoir of 1824, then to make an extensive account of it in 1849 taking advantage of the rigorous new experimental measurements made by Regnault (1847) in France.
This prompted Thomson (1848, page 102) to ask the question: ``is there a principle upon which to base an absolute thermometric scale'', before answering: yes, by using the ideas of Carnot on the motive power of fire.

This is how Thomson defined the notion of absolute temperature.
He first modified the old value of Gay-Lussac $267$ in (\ref{eq_1}) by the value $1 / 0.00366 = 273.22$ resulting from the works of Regnault (a value he approximated by $273$).
Then he explained in a note (at the bottom of page 104) that ``infinite cold must correspond to a finite number of degrees of air-thermometer below the zero '' (centigrade), since by lowering ``$t$'' reported in (\ref{eq_1}) ``we should arrive at a point corresponding to the volume of air being reduced to nothing'', a point marked by $ -273 $ degrees ``which cannot be reached at any finite temperature, however low''.

\begin{figure}[hbt]
\centering
\includegraphics[width=0.6\linewidth,angle=0,clip=true]{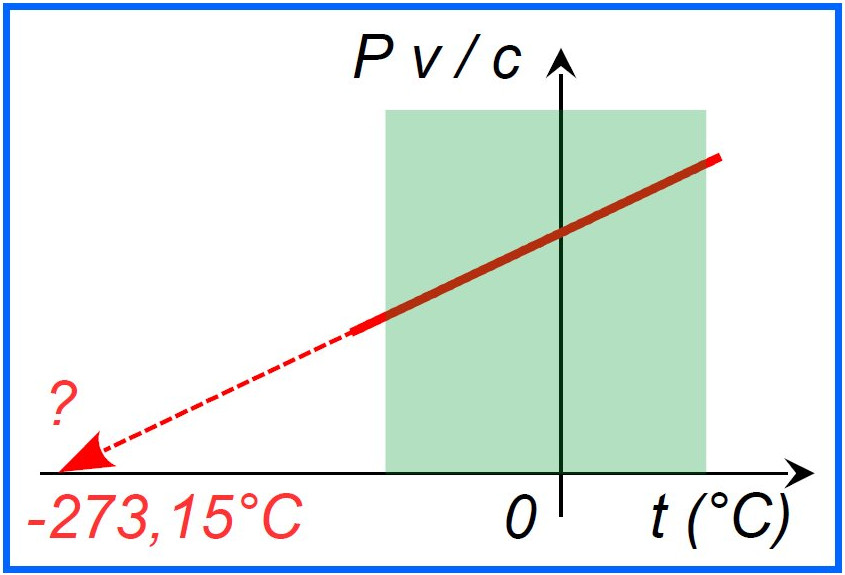}
\vspace*{-1mm}
\caption{
{\it
We draw here the quantity ``$ P \: v / c $'' given by the equation (\ref{eq_1}) as a function of the temperature ``$ t $'' in Celsius.
The usual range of atmospheric temperatures is drawn in green. 
The result is of the form ``($ t + a $)''.
It is therefore a straight line that can be prolonged (in dashed lines) up to the ordinate ``$ P \: v / c = 0$'', where the temperature was ``$ -a = -267 $''~degrees after Gay-Lussac (1802), $ - 273 $~degrees after Thomson (1848), and $ - 273.15 $~degrees in the modern Kelvin scale.
}
\label{Fig_Carnot}}
\end{figure}

This reasoning can be illustrated by Figure~1 which suggests a prolongation towards low temperatures which seems to point to a temperature around $ -273 $~degrees where, for a given pressure ($ P $), the volume ($ v $) would necessarily tend to zero.
We now use the name of William Thomson, later Lord Kelvin, to define this scale where the zero of the Celsius scale is $ 273.15 $~degrees on the Kelvin scale.
And the last words of Thomson are important, insofar as they explain that this absolute zero must be seen as an inaccessible, asymptotic limit.
We will see later the importance of this same vision for the computation of the entropy defined in an ``absolute'' way (with an inaccessible temperature of $0$~K, or asymptotically).

Does this mean that this concept of extremely low temperature and impossible to reach is easy to accept? It is quite the opposite, because one often opposes to this concept the strange and difficult side to admit that the laws which govern the changes of the gases at the usual atmospheric temperatures (say between $-100$~C and $+50$~C) can depend on something happening at $-273.15$~C\,!
  
And yet, the application of the formula (\ref{eq_1}) with the more accurate value $273$ (instead of $267$) would lead to very different results if we took another absolute origin of the temperatures, as for example the value $200$.
Thus, let us imagine that at a constant volume ($v$) we increase the temperature of a gas from $5$~C to $50$~C.
According to the formula (\ref{eq_1}) an initial pressure of $1000$~hPa would increase up to $1000$ times $323/278$, or $1162$~hPa, if one uses the Kelvin scale, or up to the slightly different value of $1165$~hPa with the old value $267$ of Gay-Lussac, but up to a much higher value of $1220$~hPa if we used the value $200$ in (\ref{eq_1}).
The experiment is easy to carry out, and it will give reason to the use of the absolute scale of the temperatures with the $273.15$ value, which thus influences in the life of every day: the physics of the usual world ``feels'' what happens at zero Kelvin, around $-273$~C.

The same ``remote'' influence is involved in the calculation of the efficiency of the thermal machines, which is equal to $1 - T_C / T_H$ and which depends on the ratio of the absolute temperatures of the systems operating between a cold source at $ T_C $ and a hot source at $ T_H $, with both $ T_C $ and $ T_H $ expressed with the Kelvin scale and not another.
And the same is true for statistical physics and for the distribution of quantized states $\exp\left[ - E / (k\:T) \right]$, which depends on the inverse of the temperature ``$ 1 / T $'', with the obligation to take for $ T $ the absolute definition of the temperature with an origin at $ -273.15 $~Celsius, excluding all other definitions.

 \section{\underline{\Large The definition of entropy}}
\label{section_3}
\vspace{-1mm}

It is again through the writings of Clapeyron, but published in German in 1843 this time, and in the knowledge of the works of Thomson, that Rudolf Clausius (1850) was able to know and use Carnot's ideas to establish a ``mechanical theory of the heat''.
Clausius then introduced in 1865 (page 390) in German and in 1867 (page 357) in English, like energy, a new function which he called ``entropy'' and which he noted ``$ S $''.
 This function is defined by Clausius by the differential equation
\vspace{-1mm}
\begin{align}  
 dS & \: = \; \frac{\delta Q}{T} \: ,
 \label{eq_2}
\end{align}
where ``$ dS $'' is an elementary evolution of entropy, $ \delta Q $ is an elementary exchange of heat that occurs during a reversible transformation, and $ T $ is the Kelvin absolute temperature introduced in the previous paragraph (excluding all other possible definitions).

Then, to know the entropy $S$ for the ambient temperature $T$, it is necessary to integrate this equation (\ref{eq_2}) between a certain temperature $T_0$ and $T$.
To do this, we must add the ``infinitely small'' $dS$ between $T_0$ and $T$, so we need to know ``$\delta Q$'' for each temperature between $T_0$ and $T$.
Moreover, the entropy $S_0$ at $T_0$ must be known. 
The knowledge of $S(T)$ is therefore only relative, since it is subordinate to that of $S_0(T_0)$ which is called ``integration constant''.
It appears that, at constant pressure $p_0 = 1000$~hPa, the quantity $\delta Q = C_p \: dT $ is proportional to the elementary evolution of the temperature ($dT$) and to the heat capacity $C_p (T)$, which depends a priori on the temperature.
To calculate the entropy at ambient temperatures ($T$) we must therefore compute the following sum\::
\vspace{-1mm}
\begin{align}  
 S(T) & \: =
 S_0 (0 \: \mbox{K}) 
 + [ \: S_1 - S_0 \: ] 
 + \frac{L_1}{T_1} 
 + [ \: S_2 - S_1 \: ]  
 + \frac{L_2}{T_2} 
 + [ \: S_3 - S_2 \: ]  
 + \frac{L_3}{T_3} 
  \: .
 \label{eq_3}
\end{align}
The quantities $ L_1 $, $ L_2 $ and $ L_3 $ are the latent heat of  change of phase (solid/solid\:\footnote{
There are 3 solid phases ($\alpha$, $\beta$, $\gamma$) and 2 solid/solid changes for $\mbox{O}_2$; 2 solid phases ($\alpha$, $\beta$) and 1 solid/solid changes for $\mbox{N}_2$. See the diagrams for $c_p(T)$ in the Fig.1 of Marquet and Geleyn (2015), available in arXiv\,: \url{https://arxiv.org/pdf/1510.03239.pdf}.},
 solid/liquid and liquid/vapour) which are divided by the respective temperatures $ T_1 $, $ T_2 $ and $  T_3$. 
The differences [$ S_1 - S_0 $], [$ S_2 - S_1 $] and [$ S_3 - S_2 $] represent the mathematical integrals of $ C_p (T) / T $ between the temperature limits $ T_0 $ and $ T_1 $, then $ T_1 $ and $ T_2 $, then $ T_2 $ and $ T_3 $.

This method of calculating the entropy of all bodies is called ``calorimetric'' because we have to integrate the values of $C_p (T) / T$, where the specific heats $C_p(T)$ must be measured with ``calorimeter'' devices.
The difficulties in calculating (\ref{eq_3}) are therefore important.
We must first know the values of the latent heats $L_1$, $L_2$ and $L_3$, also those of $C_p(T)$ for each of the temperatures between $0$~K and $T$, including for the temperatures (very) close to $0$~K.
Finally, we need to know the value of the ``integration constants'' $S_0$ at $0$~K, which must  a priori depend on each given chemical species.

 \section{\underline{\Large The importance of integration constants}}
\label{section_4}
\vspace{-1mm}

The problem of these ``integration constants'' was clearly posed for the first time by Le Chatelier (1888) in the context of the determination of chemical equilibria.
His goal was to use the characteristic functions introduced by Massieu (1869, 1876), then modified and popularized by Gibbs (1876-78) after his passage in Paris.
It was a question of calculating the variation of the free enthalpy ($G$).
It is one of the characteristic functions defined by Gibbs, whose variation induced by any chemical reaction at constant temperature is written in the form\::
\vspace{-1mm}
\begin{align}  
 \Delta G & \: = \; \Delta (H \: - T \:S\:) 
 \; = \; \Delta H \: - \: T \: \Delta S
  \: .
 \label{eq_4}
\end{align}
In (\ref{eq_4}), $H$ is the enthalpy which, for a gas, is the sum of its internal energy ($ U $) and the product of the volume ($ V $) by the pressure ($ P $), leading to $ H = U + P \: V = U \: + \: n \: R \: T $ using the ideal gas law (\ref{eq_1}).
If $\Delta G$ is to be computed for chemical reactions such as
\vspace{-1mm}
\begin{align}  
 \mbox{N}_2 \: + \: 2  \: \mbox{O}_2 
 & \: \Longleftrightarrow \; 
   2 \: \mbox{N}\mbox{O}_2
  \: ,
 \label{eq_5}
\end{align}
the symbol ``$\Delta$'' in (\ref{eq_4}) refers to the difference of the quantities $ H $ and $ S $ evaluated for the right-hand side of (\ref{eq_5}), i.e. for NO${}_{2}$, minus those evaluated for the member on the left, i.e. for N${}_{2}$ and $\mbox{O}_2$, with weighting factors depending on the stoichiometric coefficients (here: $-1$ and $-2$ for $\mbox{N}_2$ and $\mbox{O}_2$; $+2$ for $\mbox{N} \mbox{O}_2$).

Gibbs' contribution is to show that the sign of $\Delta G$ determines the equilibrium and meaning of chemical reactions such that (\ref{eq_5})~: reaction in equilibrium if ``$G$'' is minimal and if $\Delta G=0$, move to the right if $\Delta G<0$ (product NO${}_{2}$ in majority), move to the left if $\Delta G > 0$ (reactants N${}_{2}$ and O${}_{2}$ in majority).

However, since the entropy variation $\Delta S$ in (\ref{eq_4}) is multiplied by the temperature $T$, the presence of the integration constants $ S_0 (0 \: \mbox{K})$ in (\ref{eq_3}), which depends both on the reactants and the products, induce an impact which then depends on this temperature T.
And because of this, the sign of $\Delta G$ in (\ref{eq_4}) is modified by all the changes of the values of the integration constants $S_0$ which are, a priori, specific to each of the components of (\ref{eq_5}).
For this chemical reaction (\ref{eq_5}), the remaining sum in factor of $T$ writes: 
``$2 \: S_0(\mbox{N} \mbox{O}_2) - S_0(\mbox{N}_2) - 2 \: S_0(\mbox{O}_2)$''.
For this reason, the nature of the equilibrium of the chemical reactions depends on the values of $S_0$ for each of the products and reactants.
And that is why the thermochemistry tables do not treat $\Delta H$ and $\Delta S$ equivalently.
The enthalpies of reaction $\Delta H$ are directly given for all usual bodies, the integration constants for $H$ having no chemical impact.
Differently, these tables give the absolute values of the entropies $S$, which must be summed for all the products to the right of (\ref{eq_5}), then for all the reactants to the left of (\ref{eq_5}), before making the difference of these two sums and multiplying by $T$, to form the quantity ``$- \, T \: \Delta S$''.

Clausius (1865 pages 392-397, 1867 pages 359-363) and Gibbs (1876-78, pages 151-152) mentioned the existence of integration constants for energy, entropy and potentials (characteristic functions), but without noting a possible impact on physical or chemical processes.
It is Le Chatelier who first explained in 1888 in the chapter~XI entitled ``Integration constant'' (page 184) that, for entropy: ``{\it The determination of (\ldots) this constant of integration (\ldots) would bring complete knowledge of the laws of equilibrium.
It would make it possible to determine, a priori, independently of any new experimental data, the complete equilibrium conditions corresponding to a given chemical reaction\/}''.
Indeed, if we could define $\Delta S$ up to an arbitrary constant depending on each species, we could, after multiplication by $T$, modify the sign of $\Delta G$, and thus the sense of equilibrium of (\ref{eq_5}) at will, and thus remove all oxygen molecules from the atmosphere in favour of nitrogen dioxide\,!

 \section{\underline{\Large Values of the integration constants}}
\label{section_5}
\vspace{-1mm}

The first answer to the question clearly posed by Chatelier was provided by the  Nernst's ``Heat Theorem'' (``Wärmesatz'' 1906 in German, 1907 in English, ``Th\'eor\`eme de la chaleur'' 1910 in French).

\begin{figure}[hbt]
\centering
\includegraphics[width=0.99\linewidth,angle=0,clip=true]{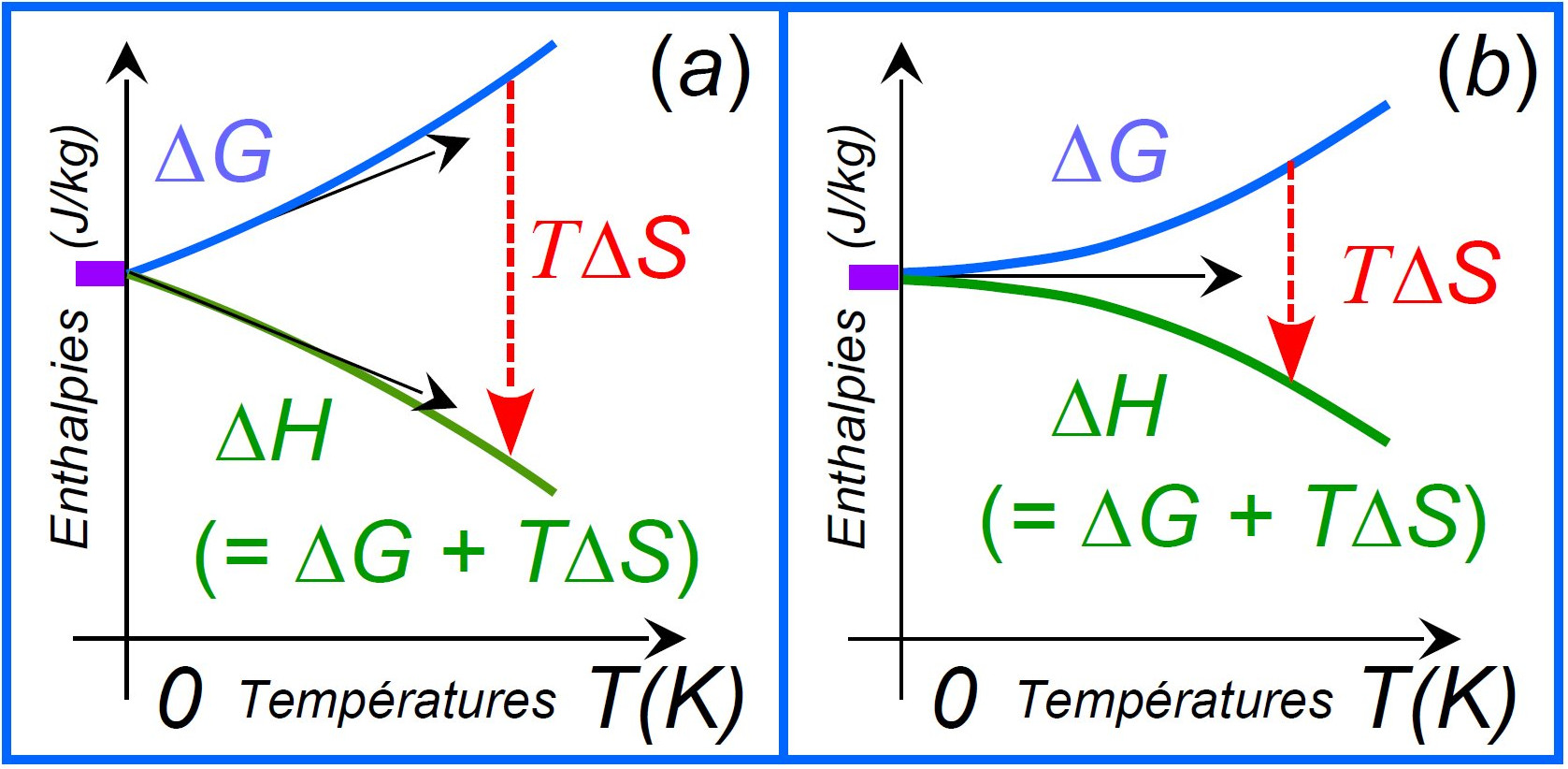}
\vspace*{-1mm}
\caption{
{\it
Graphical translation of the Nernst's theorem as expressed in his articles of 1906, 1907, 1910, but with modern notations.
The curves for the change in free enthalpy $\Delta G$ and in enthalpy $\Delta H$ are plotted in ordinate as a function of the absolute temperature on the abscissa.
}
\label{Fig_Nernst_1906}}
\end{figure}

The essence of Nernst's theorem is illustrated in Figure~\ref{Fig_Nernst_1906}. 
Experimental facts show that it is the situation described in (b) that is observed.
It appears first that $\Delta G$ and $\Delta H$ tend toward a common value for temperatures tending towards absolute zero, and therefore the product ``$T \: \Delta S$'' (the gap between the curves of $\Delta G$ and $\Delta H$) tends to zero both in (a) and (b).
The more important and new feature is that the two functions $\Delta G$ and $\Delta H$ converge towards each other like in (b) with the same horizontal tangent.
This means that the case (a) where $\Delta S$ tends toward a constant $A$ (inducing a linear variation $A \: T$ for $T \: \Delta S$) is not observed.
These facts observed by Nernst led him to state a theorem which indicates that the heat capacity $C_p$ must vary with temperature so that $\Delta S$ must tend to zero ``fast enough'' for $T$ approaching the absolute zero (namely faster than a linear law in $T$).

Einstein (1907) was able to reinforce this prediction by establishing a theoretical formula for the heat capacity $C_p$ of the solids which possessed this behavior, via a decay of the type ``$x^2 \: \exp (-x)$'' when $ 1/x $ (proportional to $T$) tends to zero (and $x$ tends to infinity).
However, this formulation did not agree with the experimental data because the decay was too fast with temperature.
This formulation has been improved by Nernst and Lindemann (1911) by making the half-sum with the first harmonic: ``$ 0.5 \: [\: x^2 \: \exp (-x) + (x / 2)^2 \: \exp (-x / 2) \:] $''.
However, this formulation was not yet fully consistent with the observations, and its introduction seemed a little bit ad-hoc.

It is precisely to better understand these disagreements on the calculations of the heat capacities $C_p$ at low temperatures close to absolute zero that the first Solvay congress was organized in 1911 (see the report by Paul Langevin and Maurice de Broglie, 1912).
It was Nernst who suggested this international meeting to Ernest Solvay, who gave him {\it carte blanche\/} to invite the greatest scientists of the time.
During this congress, the new formulation of Nernst and Lindemann was criticized by Lorentz and Einstein, the latter (page 302) also criticizing Nernst's theorem by saying: ``we can not deduce (it) from the fact that specific heats ($ C_p $) disappear near absolute zero''; even so, in his eyes: ``its legitimacy becomes more likely''.
And Einstein hypothesized that ``Nernst's theorem amounts to stating the hypothesis, moreover quite plausible, that a change (\ldots sufficiently close to absolute zero \ldots) is always possible in a purely statistic way, from the point of view of wave mechanics''.

And indeed Debye (1912) found, on the basis of reasoning using quantas and statistical physics, the correct theoretical formulation for the heat capacities $C_p$ of solids, showing (page 800) that: ``for sufficiently low temperatures the specific heat becomes proportional to the third power of the absolute temperature''.
In doing so, we can calculate via (\ref{eq_3}) the entropy of the non-metallic solids, which depends on the cube of the temperature and leading to the internal energy being in fourth power of the temperature, in agreement with the Planck's law of radiation.

Independently, Nernst (1912) responded to Einstein's criticisms by justifying his 1906 theorem differently, assuming that ``(\ldots) der absolute Nullpunkt (\ldots) nicht zu erreichen'', namely that ``the absolute zero cannot be reached'', which led to the ``principle of unattainability of the absolute zero'' clearly expressed by Simon (1927). 
We recognize the same idea expressed by Thomson (Kelvin) in 1848 on the point marked by $- 273$~degrees\:: ``which cannot be reached at any finite temperature, however low''.
This property is analogous to the new Nernst's general principle expressed in 1912, and it corresponds to an experimental truth that has never been denied until today.

\begin{figure}[hbt]
\centering
\includegraphics[width=0.99\linewidth,angle=0,clip=true]{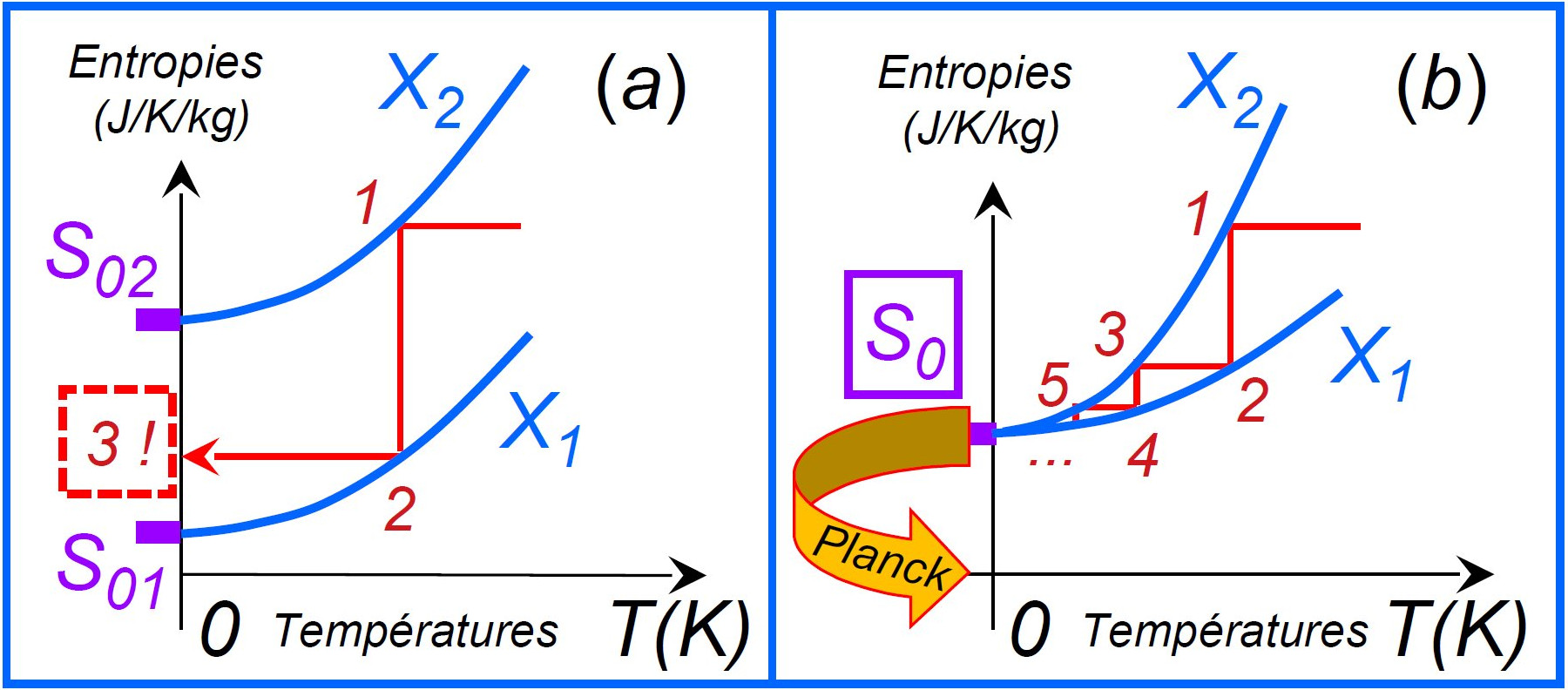}
\vspace*{-1mm}
\caption{
{\it
The entropy curves ($\:S$) are plotted as ordinates as a function of the absolute temperature ($\;T$) as abscissa.
The aim is to provide a graphical illustration of the principle of ``unattainability of absolute zero'' as expressed by Nernst in 1912 and Simon (1927).
This principle means that $S_{01}=S_{02}$ in (a).
Then, Planck (1917) postulated that this common value is a universal value $S_0$ like in (b) for the more stable solid state of all species at $T=0$~K.
The next contribution of Planck (1917) was to postulate a zero entropy $S_0 = 0$ for $T = 0$~K for all species, without loss of generality (the yellow arrow).
}
\label{Fig_Nernst_1912}}
\end{figure}

This principle can easily be understood from the two diagrams in Figure~\ref{Fig_Nernst_1912}.
We find a situation in (a) where we can reach the absolute zero in three steps: an isentrope up to the point 1 followed by an isotherm up to the point 2, then a last isentrope up to the point 3.
This situation corresponds to two formulations for the entropy of a system having an internal parameter with two distinct states $X_1$ and $X_2$ (for example magnetization), and with different entropies at zero Kelvin ($S_{01}$ different from $S_{02}$).
But only the situation described in (b) is in agreement with the experimental facts, because if the entropy $S_0$ is the same at zero Kelvin for all the parameters, then it obviously takes an infinity of steps to get closer and closer to the zero Kelvin.
We deduce that the entropy of a body cannot have several distinct values at zero Kelvin, and that $S_{01}=S_{02}=S_0$ like in (b). 
More recently, Massanes and Oppenheim (2017) consider that they have proved, in general terms, this ``principle of unattainability of absolute zero''.

 \section{\underline{\Large The third law of thermodynamics}}
\label{section_6}
\vspace{-1mm}

Max Planck (1870) passed his thesis on the second principle and the applications of entropy, less than 5 years after the paper of Clausius. 
Moreover, Planck (1900, 1901) applied Boltzmann's (1877) little-known ideas very soon, to express the formula of the black body thanks to a calculation of the entropy of the radiation written in the form\,:
\vspace{-1mm}
\begin{align}  
 S & \: = \; k_B \: \log(W) \: + \: S_0\:(0\:\mbox{K})
  \: .
 \label{eq_6}
\end{align}
In this formula (\ref{eq_6}) derived by Planck, the entropy $S$ depends on the logarithm of the number $W$ of configurations of the system, a formula that Boltzmann had suggested differently (see the English translation in Sharp and Matschinsky, 2015) and in two stages.
Boltzmann computed on the one hand the ``permutability measure'' $\Omega = \log (W) + C_1$ close to his equation (34) page 403, with an interest mentioned by Boltzmann to set $C_1 = 0$, and he computed on the other hand the change in entropy $S -S_2 = K \: \Omega + C_2$ close to his equations (62) and (65) on pages 427-428, where Boltzmann mentioned a proportionality coefficient ($K=2/3$ with certain units $R=1$ and for a monoatomic gas) corresponding to ``$k_B$'' in (\ref{eq_6}).
This formula was obtained by Boltzmann ``with a suitable choice of constant'' $C_2=0$, leading to (\ref{eq_6}) for $K=R/N=k_B$.
The entropy of Boltzmann was thus $S = K _: \ln(W) + S_2 + C_2 + K \: C_1$, and the formula of Planck (1917) corresponds to $S_0 = S_2 + C_2 + K \: C_1$.
Planck then evaluated in his two articles of 1900 and 1901 (about the black-body radiation law) both the numerical value of the constant ``$ h $'' (which bears his name) and of the constant ``$ k_B $'' of (\ref{eq_6}), giving it the name of ``Boltzmann constant'' in honour of the magnificent (but often obscure) work of Boltzmann.

In formula of Planck (\ref{eq_6}) appears the constant $ S_0 \:( 0 \: \mbox {K}) $ which has disappeared from the famous Planck formula inscribed on Boltzmann's tomb: ``$ S = k . \: \log W $''. 
This constant $S_0$ corresponds to absolute zero temperature where, by hypothesis, all the statistical microstates are degenerated into a unique configuration of maximum probability $ W = 1 $, leading to $ \log (W) = 0 $ by definition of the logarithm, and therefore to $S = 0$ if the constant $ S_0 $ is set to zero.

Planck was in fact a specialist in the computation of entropy in all areas of physics, and he was able to deliver a synthesis of his vision of thermodynamics in his 1917 treatise.
Planck added, in the third German edition, and even more in the fifth edition, a chapter on the third law where he wrote in the preface: ``The Nernst theorem in its {\it extended form\/} has in the interval received abundant confirmations and can now be regarded as well established.''

The current ``extended form'' of the third principle is the one introduced by Planck (1917, page 274) which admits, in two stages, that the absolute value of the thermodynamic entropy of any homogeneous solid at $0$~Kelvin is: 
(i) ``a universal constant''; 
(ii) ``that one can set to zero without loss of generality''. 
This second action (ii) corresponds to the yellow arrow in the Figure~\ref{Fig_Nernst_1912}(b).
Note that this choice corresponds to the cancellation of the additive constant $S_0$ in the Planck-Boltzmann statistical formula (\ref{eq_6}). 
An important caution is that this definition only concerns solids, and is therefore not applicable to perfect gases.

By a lack of luck, the water which constitutes an important component of the atmosphere possesses the singular property of having a residual entropy still present at $0$~K.
The work of Pauling (1935) completed by Nagle (1966) was used to estimate the value at $S_0 = 0.82$~cal/K/mole, by calculating the impacts due to the transfer of charges via the hydrogen bonds still at work at $0$~K within the molecule H${}_{2} $O.
So, just for water, you need to include that difference $S_0$ to have a reference that is ``the most stable'' at zero Kelvin.

As a conclusion, the third principle corresponds to setting $S=0$ for the ``more stable state'' of all solid states at $T=0$~K. 
This is equivalent to set $ S_0 \:( 0 \: \mbox {K}) = 0$ in the formula (\ref{eq_6}) of Planck,  leading to the Boltzmann's tomb formula: ``$ S = k . \: \log W $''.

It is important to note that the calorimetric values of entropy (calculated by integrating ``$ C_p (T) / T $'') are consistent with calculations made from quantum mechanics and statistical physics (computations of $W$). Indeed, the work of Tetrode and Sackur conducted from 1912 (see Grimus, 2013) resulted in a purely theoretical calculation of the entropy for monoatomic gases using the Boltzmann-Plank equation (6) with $S_0 = 0$ and by a meticulous computation of the number of complexions $W$ for the quantified energies related to the translation movements of the atoms. The same theoretical work was then carried out for the more complex molecules for polyatomic gases such as steam water (Gordon 1934), taking into account rotational and possible vibrational motions.

This equivalence can be verified by comparing the entropy measurements made with the experimental and theoretical values with reference to the table in Marquet's Appendix~A (2017), an extended version of which is given here in Table~1. 
The experimental method (calorimetric, /C) is obtained by calculating (\ref{eq_3}). 
Theoretical methods (quantum mechanics and statistical physics; /S) are based on the Boltzmann-Plank equation (\ref{eq_6}) 
with $ S_0 = 0 $ and a careful calculation of $W$ for quantized energies for translational motion, rotation or vibration, depending on the nature of the atoms or molecules.

\begin{table}
\caption{\it 
Standard molar entropies (in cal/K/mol, under one atmosphere and at $298.15$~K) for: Nitrogen (N${}_{2}$), Oxygen (O${}_{2}$), Water vapour (H${}_{2}$O), Carbon dioxide (CO${}_{2}$) and Argon (Ar).
The results are given for ST12 (Sackur, Tetrode, 1912); LG17 (Lewis and Gibson 1917); LR23 (Lewis and Randall, 1923); K32 (Kelley, 1932); G34 (Gordon, 1934); GR96 (Gokcen and Reddy, 1996) and C98 (Chase, 1998)
\label{Table}
}
\vspace*{-1mm} 
\centering
\vspace*{2mm}
\begin{tabular}{|c|c|c|c|c|c|c|c|c|}
\hline
 & ST12/S & LG17/C & LR23/C & K32/C & G34/S & GR96/C & GR96/S & C98/S \\ 
\hline
 N${}_{2}$ & & $45.59$ & $45.6$ & $45.8 \pm 0.3$ & & $45.94 \pm 0.2$ & $45.78$ & $45.796 \pm 0.005$ \\
 O${}_{2}$ & & $48.23$ & $48$ & $49.1 \pm 0.1$ & & $49.12 \pm 0.1$ & $49.02$ & $49.031 \pm 0.008$ \\
 H${}_{2}$O & & & & $51.1$ & $45.1$ & $44.31$ & $45.12$ & $45.132 \pm 0.010$ \\
 CO${}_{2}$ & & & & $50.0 \pm 2$ & & $51.13 \pm 0.1$ & $51.09$ & $51.098 \pm 0.029$ \\
 Ar & $37.0$& $36.43$ & $36.4$ & $36.8 \pm 0.2$ & & $36.96 \pm 0.2$ & $37.00$ & $37.000 \pm 0.001$ \\
\hline
\end{tabular}
\vspace*{-4mm} 
\end{table}

We see in Table~\ref{Table} that the agreement between experimental (calorimetric /C) and theoretical (quantum and statistical physics /S) values improves with time (from 1912 to 1998). 
This was largely due to measurements made by Giauque (1949) of values of $ C_p (T) $ for most known species and for values very close to absolute zero. 
This work was sanctioned by a Nobel Prize (see Tiselius, 1949). 
Giauque insists on the agreement of his measurements with the third principle, which together, as imagined by Le Chatelier in 1888, make it possible to predict the stability regimes of all chemical reactions.

It may be mentioned that the third principle is only partially used in some official definitions of thermochemical tables used in atmospheric and oceanic studies, such as those of IAPWS (2010) and Feistel et al. (2010), with however an explicit mention of the possibility to use the third law expressed in Feistel and Wagner (2006) for the water Ice-Ih, and also by Lemmon et al. (2000) who used a third-value for the reference entropy of the dry air formed of a mixture of N${}_{2}$, O${}_{2}$ and Ar.\:\footnote{
It is easy to modify the TEOS-10 software in order to compute the absolute entropy of moist air based on the third law of thermodynamics.
The water vapour entropy is set in ``\texttt{init\_iapws95}'' in ``\texttt{Flu\_1.F90}'', where the present value $n^0_1=-8.32044648374969$ can be replaced by $-15.94$, with a difference given by the ratio $- s^0_l/R_v \approx -3517/461.52 \approx -7.62$. 
The dry air entropy is set in ``\texttt{init\_iapws10}'' in ``\texttt{Air\_1.F90}'', where the present value $n^0_4=9.7450251743948$ can be replaced by $-13.835$ (which very close to the value published in Lemmon et al., 2000). 
The difference from the IAPWS value is given by the ratio $-s^0_d/R_d \approx -6770/287.12 \approx -23.58$.
}
It is therefore easy to modify two lines in the software TEOS-10 of the IAPWS to agree with the third law.

Some of the values of the entropies listed in Table~1 have been recalculated in Marquet (2015), the aim being to validate the experimental values used to calculate the thermal enthalpy with respect to absolute zero. 
The agreement was correct for the entropies of N${}_{2}$ ($ 46.0 \pm 0.2 $) and H${}_{2}$O ($ 45.2 \pm 0.1 $), with a larger value for O${}_{2}$ ($ 49.7 \pm 0.4 $).
This larger value is obtained by taking into account the second order transition forming a kind of Dirac peak for the $ C_p $ of O${}_{2}$ in solid phase at $ 23.85 $~K. 
I was able to observe on this occasion that the biggest error in the value of the entropy comes from the terms depending on the latent heats of phase change, and much less of the integrals of $ C_p (T) $ from $ 0 $~K to $273.15$~K.

Finally, let us recall the Schrödinger's synthesis (1944, pages 15-17, ``Discussion of the Nernst theorem'') read during his seminars in Dublin where he taught courses in statistical physics:
i) the important thing for this integration constant (i.e. $S_0$) is not the zero value set by Planck;
ii) the important thing is that it is a quantity that is independent of all the physical parameters (temperature, pressure, magnetism, electric charge, \ldots) and also of the nature of the bodies in general;
iii) this constant is, in the end, independent of all the bodies and it is a quantity that we can set to zero without loss of generality, as Planck did, just to avoid any confusion and temptation to choose another possibly variable and inappropriate value.

 \section{\underline{\Large Conclusions}}
\label{section_7}
\vspace{-1mm}

Mistrust of the validity of the third principle of thermodynamics continues. 
This is the reason why it seemed important to recall why the greatest scientists, such as Nernst, Einstein and Planck as early as 1911, till Schrödinger in 1944, considered this scientific principle as verified by all the experimental facts and in agreement with the best theoretical developments based on statistical physics and quantum mechanics.

This article will form the basis for a second part (Marquet, 2019) which will present the different ways of calculating the moist-air entropy of the atmosphere consisting of a mixture with variable proportions of dry air, water vapour, cloud liquid water, cloud ice and liquid or solid precipitations.

It will be seen in the second part that the same questions and doubts about the validity of the third law, with a possible influence of a hypothesis made at the absolute zero of temperatures, are at work in meteorology and in the climate sciences.
However, the entropy defined with the third law corresponds to new, singular and very interesting properties, and should be taken into account to better understand certain properties of the moist-air atmosphere consisting of a gas with a variable quantity of water vapour.

 \section{\underline{\Large References}}
\label{section_Biblio}
\vspace{-3mm}

\noindent
$\bullet$
Boltzmann, L., (1877). Über die Beziehung zwischen dem zweiten Hauptsatz der mechanischen Wärmetheorie und der Wahrscheinlichkeitsrechnung, respective den Sätzen über das Wärmegleichgewicht (On the relation between the second law, the mechanical theory of heat, the theory of probability, and the theorems of heat balance). 
{\it Sitzb. d. Kaiserlichen Akademie der Wissenschaften, mathematisch naturwissen\/}. 
Vol. 76, Number 3, 373--435.

\noindent
$\bullet$
Carnot,  S., (1824). 
{\it Réflexions sur la puissance motrice du feu (Reflections on the motive power of fire)\/}. 
Bachelier, Paris,  118~p.

\noindent
$\bullet$
Chase, M.W., Jr., (1998). 
{\it JANAF Thermochemical Tables. 4th ed.\/},
American Chemical Society, 1951~p.

\noindent
$\bullet$
Clapeyron, E., (1834). 
Mémoire sur la puissance motrice de la chaleur (Memory on the motive power of heat). 
{\it Journal de l'école royale polytechnique\/}. 
Cahier 22, Tome 14, 153--190.

\noindent
$\bullet$
Clapeyron, E., (1837). 
{\it Memoir on the Motive Power of Heat\/}. 
Scientific Memoirs. Edited by  Richard Taylor, 
Volume 1, Part 3, Article 15, 347--376.

\noindent
$\bullet$
Clapeyron, E., (1843). 
Ueber die bewegende Kraft der Wärme (About the motive power of heat). 
{\it Annalen der Physik und Chemie\/}.  
59, 2nd serie, 446--451 et 566--586.

\noindent
$\bullet$
Clausius, R., (1850). 
Ueber die bewe bewegende Kraft der Wärme und die Gesetze, whelche sich daraus für die Wärmelehre selbst albeiten lassen
(On the moving force of heat and the laws which can be worked out for the theory of heat itself).  
{\it Annalen der Physik und Chemie\/}, 
79, 368--397 and 500--524.

\noindent
$\bullet$
Clausius, R. (1865). 
Ueber verschiedene für die Anwendung bequeme Formen der Hauptgleichungen der mechanischen Wärmetheorie 
(On various convenient forms of the main equations of mechanical theory of heat). 
{\it Annalen der Physik und Chemie\/}, 
125, 353--400.

\noindent
$\bullet$
Clausius, R. (1867). 
{\it On several convenient forms of the fundamental equations of the mechanical theory of heat\/}, 327-374. 
{\it The mechanical theory of heat\/}.  
J. Van Voorst, London, 374~p.

\noindent
$\bullet$
Debye, P. (1912). 
Zur Theorie der spezifischen Wärmen (The theory of specific heat).
{\it Annalen der Physik\/}, 
39, 789--839.

\noindent
$\bullet$
Einstein, A. (1907).
Die Plancksche Theorie der Strahlung und die Theorie der spezifischen Wärme. 
{\it Annalen der Physik\/}, 327: 180--190.
An English translation ``Planck's theory of radiation and the theory of specific heat'' is available in the Doc.38 (p.214-224) in {\it The Collected Papers of Albert Einstein, Volume 2: Writings 1900-1909 (English translation supplement)\/}.
\url{https://einsteinpapers.press.princeton.edu/vol2-trans/228}

\noindent
$\bullet$
Feistel, R., Wagner, W., (2006). 
A new equation of state for H2O ice Ih.  
{\it J. Phys. Chem. Ref. Data\/}, 
35, 1021--1047.

\noindent
$\bullet$
Feistel, R.,  Wright, D.G.,  Kretzschmar, H.J.,  Hagen, E.,  Herrmann, S.,  Span, R., (2010).  
Thermodynamic Properties of Sea Air. 
{\it Ocean Sci.\/}, 
6,  91--141.

\noindent
$\bullet$
Gay-Lussac, L.F., (1802). 
Recherches sur la dilatation des gaz et des vapeurs (Research on the expansion of gases and vapours). 
{\it Annales de Chimie\/}, 
43, 137--175.

\noindent
$\bullet$
Giauque, W.F., (1949).  
{\it Some consequences of low temperature Research in chemical thermodynamics\/}. 
Nobel lectures, chemistry 1942-1962. Elsevier Publishing Company, Amsterdam, 1964.  227--250.

\noindent
$\bullet$
Gibbs, J.W., (1876-1878). 
{\it On the equilibrium of heterogeneous substances\/}. 
p.108--248 and p.343--520. 
Transactions of the Connecticut Academy of Art and Sciences, 
Vol. III, New Haven.

\noindent
$\bullet$
Gokcen, N.A., Reddy, R.G., (1996). 
{\it Thermodynamics (Chapter VII on the Third Law)\/}.  
Springer Science+Business Media, New York, 400~p.

\noindent
$\bullet$
Gordon, A.R., (1934).  
The Calculation of Thermodynamic Quantities from Spectroscopic Data for Polyatomic Molecules; the Free Energy, Entropy and Heat Capacity of Steam. 
{\it J. Chem. Phys.\/}, 
2,  65-72.

\noindent
$\bullet$
Grimus, W., (2013). 
100th anniversary of the Sackur-Tetrode equation. 
{\it Ann. Phys. Berlin\/}, 525, A32-A35. 
\url{https://arxiv.org/abs/1112.3748}

\noindent
$\bullet$
IAPWS G8-10,  (2010).  
{\it Guideline on an equation of state for humid air in contact with seawater and ice, consistent with the IAPWS formulation 2008 for the thermodynamic properties of seawater\/}.
International Association for the Properties of Water and Steam. 21~p. 
\url{http://www.iapws.org/relguide/SeaAir.html}

\noindent
$\bullet$
Kelley K.K., (1932). 
{\it Contributions to the Data on Theoretical Metallurgy: 
I. The entropies of inorganic substances\/}. 
Bulletin~350. 
United States, Government Printing Office, Washington. 63~p.

\noindent
$\bullet$
Langevin, P., de Broglie, M., (1912). 
{\it La théorie du rayonnement et les quanta. Rapports et discussions de la réunion tenue à Bruxelles, du 30 octobre au 3 novembre 1911 (The theory of radiation and quanta. Reports and discussions of the meeting held in Brussels from 30 October to 3 November 1911)\/}. 
Gauthier-Villars, Paris. 
461~p.

\noindent
$\bullet$
Lemmon, E. W., Jacobsen, R. T., Penoncello, S. G., and Friend, D.G., (2000). 
Thermodynamic properties of air and mixtures of Nitrogen, Argon and Oxygen from 60 to 2000 K at pressures to 2000 MPa.
{\it J. Phys. Chem. Ref. Data\/}, 29, 331--385.

\noindent
$\bullet$
Lewis, G.N., Gibson, G.E., (1917). 
The entropy of the elements and the third law of thermodynamics. 
{\it J. Am. Chem. Soc.\/}, 39 , 2554--2581.

\noindent
$\bullet$
Lewis, G.N., Randall, M., (1923). 
{\it Thermodynamics and the Free Energy of Chemical Substances\/}. 
McGraw-Hill Book Company, New-York, 653~p.

\noindent
$\bullet$
Le Chatelier, M.H., (1888). 
{\it Recherches expérimentales et théoriques sur les équilibres chimiques (Experimental and theoretical researches on chemical equilibria)\/}. 
Dunod Editeur, Paris. 
229~p.

\noindent
$\bullet$
Marquet, P., (2015). 
On the computation of moist-air specific thermal enthalpy. 
{\it Quart. J. Roy. Meteorol. Soc.\/}, 
141,  67--84.  
\url{http://arxiv.org/abs/1401.3125}

\noindent
$\bullet$
Marquet, P., (2017). 
A Third-Law Isentropic Analysis of a Simulated Hurricane. 
{\it J. Atmos. Sci.\/}, 
74, 3451--3471. 
\url{https://arxiv.org/abs/1704.06098}

\noindent
$\bullet$ Marquet, P., (2019). 
Le troisi\`eme principe ou une definition absolue de l'entropie. 
Partie~2~: ses d\'efinitions et applications en M\'et\'eorologie et en Climat. 
{\it La M\'et\'eorologie\/}, Accepted for publication.
\url{http://documents.irevues.inist.fr/handle/2042/14834}
The third law of thermodynamics or an absolute definition for Entropy. 
Part~2: definitions and applications in Meteorology and Climate.
\url{https://arxiv.org/abs/1904.11699}

\noindent
$\bullet$
Masanes L.,  Oppenheim J., (2017). 
A general derivation and quantification of the third law of thermodynamics.
{\it Nature Communications\/}. 
8, 1--7.

\noindent
$\bullet$
Massieu F., (1869). Sur les fonctions caractéristiques des divers fluides (On the characteristic functions of the various fluids). {\it Comptes rendus de l’Académie des Sciences de Paris\/}, 69, 858-862 et 1057-1061.

\noindent
$\bullet$
Massieu F., (1876). Mémoire sur les fonctions caractéristiques des divers fluides et sur la théorie des vapeurs (Memoir on the characteristic functions of the various fluids and on the theory of vapors). {\it Mémoires pour l’Académie des Sciences de l’Institut de France\/}, Paris, 22, 1-92.

\noindent
$\bullet$
Nagle, J.F., (1966).  
Lattice statistics of Hydrogen bonded crystals. I. The residual entropy of Ice. 
{\it J. Math. Phys.\/}, 
7, 1484--1491.

\noindent
$\bullet$
Nernst, W., (1906). 
Ueber die Berechnung chemischer Gleichgewichte aus thermischen Messungen 
(On the calculation of chemical equilibria from thermal measurements). 
{\it Nachrichten von der Gesellschaft der Wissenschaften zu Göttingen, 
Mathematisch-Physikalische Klasse\/},
1--40. 

\noindent
$\bullet$
Nernst, W., (1907). 
{\it Experimental and theoretical applications of thermodynamics to chemistry. 
Lectures at the Yale University (in 1906)\/}. 
Charles Scribner's Sons, New York, 39--52.

\noindent
$\bullet$
Nernst, W., (1910). 
Sur les chaleurs spécifiques aux basses températures et le développement de la thermodynamique (On the specific heats at low temperatures and the development of thermodynamics). 
{\it J. Phys. Theor. Appl.\/} 
9, 721--749.

\noindent
$\bullet$
Nernst, W., (1912).  
Thermodynamik und spezifische Wärme (Thermodynamics and specific heat). 
{\it Preussische Akademie der Wissenschaften (Berlin)\/}. 
Sitzungsberichte, 
Jan-Jun, 134--140.

\noindent
$\bullet$
Nernst, W., Lindemann, F.A.,  (1911).  
Untersuchungen über die spezifische Wärme bei tiefen Temperaturen 
(Investigations on the specific heat at low temperatures). 
{\it Preussische Akademie der Wissenschaften (Berlin)\/}. 
Sitzungsberichte, 
20, 494--501.

\noindent
$\bullet$
Pauling, L., (1935).  
The structure and entropy of ice and of other crystals with some randomness of atomic arrangement. 
{\it J. Am. Chem. Soc.\/}, 
57,  2680--2684.

\noindent
$\bullet$
Planck, M., (1870). 
{\it Über den zweiten Hauptsatz der mechanischen Wärmetheorie 
(About the second law of mechanical theory of heat)\/}. 
Theoror Ackermann, 
Munich. 
61~p. 

\noindent
$\bullet$
Planck, M., (1900).  
Zur Theorie des Gesetzes der Energieverteilung im Normalspectrum (On the theory of the Law of Distribution of Energy in the normal spectrum).  
{\it Verhandlungen der Deutschen Physikalischen Gesellschaft\/}. 
2, 237--245

\noindent
$\bullet$
Planck, M., (1901). 
Über das Gesetz der Energieverteilung im Normalspektrum (About the law of  distribution of energy in the normal spectrum). 
{\it Annalen der Physik\/}. 
4, 553--563.

\noindent
$\bullet$
Planck, M., (1917). 
{\it Treatise on Thermodynamics\/}. 
Translated into English by A. Ogg from the seventh German edition. 
Dover Publication, Inc., 297~p.

\noindent
$\bullet$
Regnault V., (1847). 
{\it Relation des expériences (\ldots) pour déterminer les
principales lois et les données numériques qui entre 
dans le calcul des machines à vapeur
(Relationship of experiments (\ldots) to determine the
major laws and the numerical data that enters
in the calculation of steam engines)\/}. 
Firmin Didot Frères, Paris, 767~p.

\noindent
$\bullet$
Schrödinger E., (1952). 
{\it Statistical thermodynamics\/}. 
A course of seminar lectures delivered in January-March 1944, 
at the School of theorical Physics, 
Dublin Institute for Advanced Studies. 
Cambridge University Press, 95~p.

\noindent
$\bullet$
Sharp, K.,  Matschinsky, F., (2015).
Translation of Ludwig Boltzmann's Paper ``{\it On the Relationship between the Second Fundamental Theorem of the Mechanical Theory of Heat and Probability Calculations Regarding the Conditions for Thermal Equilibrium\/}'' (published in the Sitzungberichte der Kaiserlichen Akademie der Wissenschaften, Mathematisch-Naturwissen Classe, Abt. II, LXXVI 1877, pp 373-435).
{\it Entropy\/} 17 1971--2009.

\noindent
$\bullet$
Simon, F. (1927). 
Zum Prinzip von der Unerreichbarkeit des absoluten Nullpunktes 
(On the principle of the unattainability of absolute zero).
{\it Zeitschrift für Physik\/}, 41, 806--809.

\noindent
$\bullet$
Tiselius, A.,  (1949),  
{\it The Nobel prize in chemistry 1949. Award ceremony speech\/}.  
Nobel Lectures, Chemistry 1942--1962.  
Elsevier Publishing Company, Amsterdam, 1964.

\noindent
$\bullet$
Thomson, W., (1848). 
On an absolute thermometric scale founded on 
Carnot's theory of the motive power of heat, 
and calculated from Regnault's observations. 
Philosophical Magazine. 
Reprinted in {\it Mathematical and Physical papers of William Thomson\/}, 
Article 39, Vol. 1, 100--106, 1882.

\noindent
$\bullet$
Thomson, W., (1849).
An account of Carnot's theory of the motive power of heat, 
with numerical results deduced from Regnault's experiments on steam. 
Transactions of the Edinburgh Royal Society. 
Reprinted in {\it Mathematical and Physical papers of William Thomson\/},
Vol. 1, Article 41, 113--155, 1882.

  \end{document}